# Implementation of Data Mining on a Secure Cloud Computing over a Web API using Supervised Machine Learning Algorithm

Data Mining in a Secure Cloud Computing Environment through Restful API

Tosin Ige[1]
Department of Computer Science
University of Texas at El Paso
Texas, USA

Sikiru Adewale[2]
Department of Computer Science
Virginia Technological University
Virginia, USA

*Abstract*—Ever since the era of internet had ushered in cloud computing, there had been increase in the demand for the unlimited data available through cloud computing for data analysis, pattern recognition and technology advancement. With this also bring the problem of scalability, efficiency and security threat. This research paper focuses on how data can be dynamically mine in real time for pattern detection in a secure cloud computing environment using combination of decision tree algorithm and Random Forest over a restful Application Programming Interface (API). We are able to successfully Implement data mining on cloud computing bypassing or avoiding direct interaction with data warehouse and without any terminal involve by using combination of IBM Cloud storage facility, Amazing Web Service, Application Programming Interface and Window service along with a decision tree and Random Forest algorithm for our classifier. We were able to successfully bypass direct connection with the data warehouse and cloud terminal with 94% accuracy in our model.

I. INTRODUCTION

As we all know that Knowledge discovery is the nontrivial extraction of implicit, previously unknown, and potentially useful information from data [1]. There is no doubt that the availability of billions of data in the cloud had open floodgates of opportunity in which model can be trained and learn by itself over time thereby enable machines and intelligent elements to make crucial and important decision without human intervention. As millions of data are constantly being stored and retrieve through cloud computing on daily bases. There is no doubt that data mining delivers a powerful competitive advantage [6] for any company or industry. With these also arises the challenge on how to make use of the unlimited data available through data mining on cloud computing, and how it can be done in a highly secure, scalable and efficient manner to combat several security threats associated with cloud computing.

In modern research and intelligent systems, the important of data can never be overemphasized across wide range of industries, especially when we consider the fact that the traditional E-commerce businesses and industry were influenced by cloud computing in technical architecture, service modes and the industrial chain [4], economist needs to know how they can use readily available data in the cloud to predict consumers need and behavior, meteorologists need those data to make future weather forecast and predict or detect climatic change, government agency needs those data to make effective policy, police and intelligence agency needs data for background check and so on. All government and non-government parastatal depends on data from the cloud for one reason or the other. This further brings about the importance and urgency of mining data to detect previously unknown pattern in a scalable and efficient manner through cloud computing to address daily needs with the guarantee of maximum security, authorization and authentication which can be affected through a restful web API.

In this research, the use of constantly updating real time data and without any connection to the cloud data warehouse for maximal data protection and dynamic predictive pattern with high accuracy were implemented. To achieve this objective, an application Programming Interface (API) and Background window service was developed to detect and fetch new and updated records from data warehouse, transformed to json and written to a file on the cloud. This ensures maximum security to the data as there is no direct connection to the data warehouse where data are constantly being pulled from. We used IBM Cloud object storage facility to host our csv file while the data warehouse is also on the cloud but from another channel entirely. Then a restful web API service was developed using Django python framework which was deployed to Amazon Web Services (AWS). Since the newly created API will be constantly pulling data from the data warehouse and writing directly to into the CSV file which is on the IBM Cloud storage facility. The service can easily be overwhelmed, to prevent the service from being overwhelm, we developed a background service on a Microsoft console using Microsoft .NET Technology programming language.

Any record the web API misses or delayed in picking from the warehouse, the background service will pick it up. Hence, they complement each other and ensure efficiency so that the web API service is not overwhelmed. With this in place, there is a constantly updated and up to date data on the CSV file hosted in the IBM cloud storage facility which is now serves as the primary source of data. This ensure accessibility to updated records at every stage of our data mining activities, as there is an existing restful API and background window service that fetches data from the warehouse to the hosted file in cloud environment from which data is being pulled from.

Combination of decision tree algorithm and Random Forest for classifier was implemented coupled with a graphic user interface (GUI) that automatically processed and displays newly discovered patterns in a graphical format on the screen at each and every update on new data.

There are four main objectives here which are:

*1)* Bypassing direct connection and retrieval from the data warehouse through a middle ware.

*2)* Real time access to dynamic data in an enabling cloud environment.

*3)* We want to shield the warehouse for maximum security and while also avoiding data lock.

*4)* Automation of the detection of any new pattern from the dataset and projection to the screen with graphical illustration and analysis.

To ensure successful accomplishment of the fourth objective, a scheduler called Python scheduling library was used so that the whole process is automatically initiated and repeated immediately a new pattern is detected.

## II. BACKGROUND STUDY

The importance of cloud computing cannot be overemphasize as it includes; unlimited storage, provisioning and updating, guaranteed privacy more security [13]. Also, it is possible for users of cloud services to optimize server utilization, dynamic scalability, and minimize the development of new application life cycles [14]. In addition to the numerous benefits in cloud computing, there are also problems as a result of cloud outage since data storage is centralize in the cloud which can paralyze a company business [15], also attack on the integrated cloud environment can cause loss of data and finance for both the service providers and subscribers. There are also other risks associated with computing on cloud environment which includes the issues of threats to data security information confidentiality [5], and the possibility of information leakage and vulnerability [12].

In today's data mining, multiple data streams are generally complex than single data streams [2] considering the security architecture and cloud computing environment, the complexity increases when we consider the cost and benefit, reliability, cloud migration and inter clouding [3], While compute/storage scaling, data parallelism, virtualization, MapReduce, RIA, SaaS and Mashups are likewise also important in data mining [7] not all are always implemented. Although there had been improvement in the technology and services of the major Cloud Computing Platforms likes amazon Relational Database Service, Amazon Simple Queue Service, Amazon SimpleDB, Amazon Web Services, AppScale, Azure Services Platform, Caspio, CloudControl, Cordys Process Factory, Engine Yard, Force.com, FreedomBox, Google App Engine, Heroku, Hybrid Web Cluster, OrangeScape, Platform as a service, Rackspace Cloud, Rollbase, Squarespace, Sun Cloud, Vertebra (cloud computing framework), Wolf Frameworks [8], some of the persistent problems associated with data mining on cloud computing are as a result of the existing method adopted in the industry and scientific world at large. Current cloud computation for data mining needed service provider to provide interface for user. The user does not need to bother about the infrastructure. It enables the retrieval of useful previously unknown data from integrated data warehouse, in such a way that users doesn't need to border about infrastructure, storage, or configuration and maintenance, the provider handles that. It is based on retrieving directly from the integrated data warehouse.

[5] discusses the importance of large item sets within a cluster and its implication for effectiveness, it doesn't address the several issues and vulnerabilities in cloud computing while [9] in his research work "Data mining in Cloud Computing", relies on extraction of previously unknown or meaningful pattern from unstructured or semi-structured data from the web sources; "The analysis steps in the Knowledge Discovery and Databases process" [11]. It listed three stages of research involving data warehouse which are staging, integration, and accessibility for the purpose of reporting and analysis in the Review of Data Mining Techniques in Cloud Computing Database by [10].

In addition to the problems associated with data mining on cloud computing in which majority of the problems are due to existing methods of data mining on cloud computing. The existing methods also creates a wide gap between data mining on cloud computing and application Programming Interface (API) which are yet to be closed. In the existing method we are yet to see an instance in which we call we only need to call an API endpoint, and then have everything done for us.

## III. RESEARCH METHODOLOGY

Data used on this research work was obtained from social network on GitHub. To begin this research, four basic things were paramount;

*1)* Subscription to IBM Cloud Object Storage facility to host our CSV file.

*2)* Setting up a data warehouse on cloud from another channel different from IBM.

*3)* Development of a background window service using .NET Technology.

*4)* Implementation of a web service to be consumed in the cloud using python.

*5)* Scheduler using python scheduler library to trigger the web API at interval.

In order to prevent the web service from being overwhelm due to multiple calling of the endpoint at regular interval, we developed a background window service to sup- port the restful API service. Both the web API and the window service are picking records from the integrated data warehouse and pushing to the CSV file on the IBM Object Cloud Storage facility. They automatically pick new records to the CSV, and if a record is modified, it will be picked and modified on the CSV as well. The essence of the scheduler which was written in python is to be calling the web API at regular interval to check and push from the integrated data warehouse to the CSV file.

With the successful setting up of cloud environment and the necessary software programs being in fully execution, we proceeded by adopting the following data cleansing and preparation techniques;

Data cleaning: We use preprocessing and cleaning methods to remove incomplete data that might cause system failure and also affect output prediction. Rows containing missing values where completely removed. We also use different methods to identify and remove noisy data, outliners, and other factors which can influence the output result.

Data Reduction for Data Quality: In order to maintain data integrity, we needed to deal with all rows containing null value; hence we opted for python library tool called PyCaret. We have two options which are either to automatically fill all the null values or to remove any row(s) containing null values weighted. Having weighted the risk involves in both, we decided to remove any row with null or empty value from the data, and this was done using PyCaret python library.

Data Transformation: For us to make our data to acceptable format for easy data mining and pattern recognition, it needed to undergo some data transformation. To ensure data is fully transformed to acceptable format, we used Discretization, normalization, and data aggregation technique.

Data Mining: Having successfully set up our apparatus which includes IBM cloud object storage facility, running background window service, active web service, scheduler, and with the data being thoroughly pre-processed and transformed.

Unlike current data mining in cloud computing process in which, one needs to make direct connect to the data warehouse or directly call the csv file. We only called our restful API endpoint (Fig. 1) which was developed and hosted on the cloud.

This ensures a higher level of security and control over the data, the only thing that needed to be called is the endpoint of our API which automatically displays the data as seen in Fig. 1 displaying the first five (5) records in the data using python library in panda. (Fig. 2) shows the text representation of the selected features using decision tree algorithm.

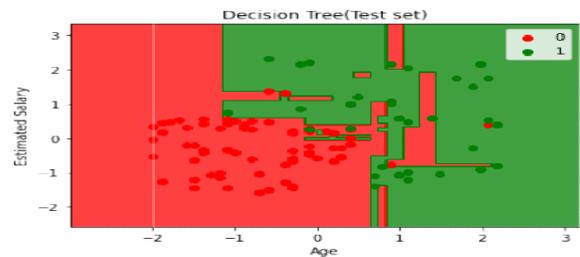

Fig. 1. Preview of First Five Rows of Dataset.

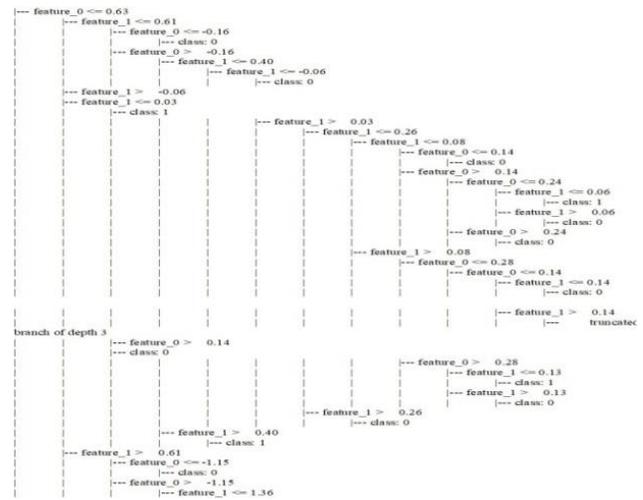

Fig. 2. Text Representation of the Features using Decision Tree.

Optimized Decision tree algorithm was used to train the model having slatted the data into two equal part, half of it to train the model and the remaining half for testing of the model as seen in (Fig. 3).

```
feature_cols = ['Age','EstimatedSalary' ]
X = data.iloc[:,[2,3]].values y = data.iloc[:,4].values
#split the dataset into training and test
X_train, X_test, y_train, y_test = train_test_split(X,y,test_size = 0.25, ran-
dom_state= 0) #perform feature scaling
from sklearn.preprocessing import StandardScaler sc_X = StandardScaler()
X_train = sc_X.fit_transform(X_train) X_test = sc_X.transform(X_test)
#fit the model in the decision tree classifier from sklearn.tree import Decision-
TreeClassifier classifier = DecisionTreeClassifier()
classifier = classifier.fit(X_train,y_train)
```

Fig. 3. Code Snippet for Important Feature Selection and Training of Model.

We are able to obtain 90% accuracy on testing our model, after which we decide to optimize for more accuracy and better performance as seen in Fig. 4.

Fig. 4. Model Accuracy and Test Performance Evaluation.

We are able to obtain accuracy of 94% after which we decided to visualize the performance of the model.

We needed to know the level of fitting, because having an accuracy of 94% can be as a result of over fitting of the model with the training set of data, hence we decided to do two additional things. Firstly, we implemented Random Forest Algorithm for training the data again to check performance and then optimize, it gives an accuracy of 92%. Secondly, we fed another set of data in the format of the trained data but new to the model, the model performed very well with accuracy. So, we proceeded to measure the performance of the model using confusion matrix, classification report, and accuracy score which gives good indication of optimal performance. We are in a dilemma either to ensure about 100% accuracy or over fitting because in supervise machine learning, high accuracy of almost 100% can be as a result of over fitting of the model which we want to avoid by possible means. So, since we are able to feed the model with new set of data which had not been previously fed to the model for which it performed very well with high rate of accuracy. So, we gave priority to avoid over fitting of the model than the accuracy of the model since high accuracy for supervised learning can be as a result of over fitting of the model for which the model becomes less accurate or behaved weird when fed with unfamiliar data. The validation report can be seen in Fig. 5.

```
Mean Absolute Error: 0.09166666666666666
Mean Squared Error: 0.09166666666666666
Root Mean Squared Error: 0.30276503540974914

[[61  4]
 [ 7 48]]

              precision    recall  f1-score   support

           0       0.90      0.94      0.92        65
           1       0.92      0.87      0.90        55

    accuracy                           0.91       120
   macro avg       0.91      0.91      0.91       120
weighted avg       0.91      0.91      0.91       120
```

Fig. 5. Validation, Cross Validation and the Estimation of Mean Square Error (MSE).

## IV. CONCLUSION

In this applied research, we are able to achieve four goals:

*1)* Avoid direct interaction with integrated data warehouse.

*2)* We are able to access data in a secure cloud computing environment.

*3)* We are able to close the existing gap in data mining and web API, as all we needed to call is the endpoint of our web API. No need of terminal, or uniform resource locator (URL) of any csv is involved. It is simply over a web API.

*4)* As the patterns changes in the background, the algorithm automatically adjusts with accuracy of ninety-four (94) percent after optimization.

We are able to implement data mining in a secure cloud computing environment with accuracy of ninety-four (94) percent after optimization in our decision tree algorithm over web API. All that an AI engineer, data scientist, or machine learning engineer needs is just the endpoint of the API. This removes complexity while at the same time simplifying the whole process, it also add additional layer of security, and also remove unnecessary bottleneck as scientist and engineers will be able to concentrate more on optimizing their algorithm and model for optimal result since only API endpoint is what is needed to be called.

We hope that this will be de-facto standard in the data mining, machine learning, data science and other similar industry at large.

## V. LIMITATION AND FUTURE RESEARCH WORK

This research work is based on the quantity of data available to us. Also, as Infrastructure As a Service Provider (IAAS), Software As a Service Provider (SAAS), and Platform As a Service Provider (PAAS) continually improves their service for more secured and enhanced cloud computing environment, over the times, this can affect the performance of the model and the overall architecture, also as more data becomes available, there is possibility of more false positive and this directly impacts the bias-variance tradeoff.

So, there is future research work of developing what we called intelligent model, model which can detect availability or changes in data, detect changes in the cloud computing environment and then re-trained and re-adjust itself over the time in line with those changes.